\documentclass[final]{raa06}           
\usepackage{graphicx,times}             
\usepackage{natbib}
\usepackage{amssymb,amsmath}
\bibpunct{(}{)}{;}{a}{}{,}
\usepackage[colorlinks=true, citecolor=blue]{hyperref}%

\newcommand{\cm}{{~\rm cm}}

\newcommand{\km}{{~\rm km}}
\newcommand{\s}{{~\rm s}}

\newcommand{\g}{{~\rm g}}
\newcommand{\G}{{~\rm G}}
\newcommand{\K}{{~\rm K}}

\newcommand{\yr}{{~\rm yr}}

\begin{document}

   \title{Pre-explosion helium shell flash in type Ia supernovae
}

   \volnopage{Vol.0 (20xx) No.0, 000--000}      
   \setcounter{page}{1}          

   \author{Noam Soker
      \inst{1}
   }

   \institute{Department of Physics, Technion, Haifa, 3200003, Israel;   {\it soker@physics.technion.ac.il}\\
\vs\no
   {\small Received~~20xx month day; accepted~~20xx~~month day}}

\abstract{
I study the possibility that within the frame of the core degenerate (CD) scenario for type Ia supernovae (SNe Ia) the merger process of the core of the asymptotic giant branch (AGB) star and the white dwarf (WD) maintains an envelope mass of $\approx 0.03 M_\odot$ that causes a later helium shell flash. I estimate the number of pre-explosion helium shell flash events to be less than few per cents of all CD scenario SNe Ia. A helium shell flash while the star moves to the left on the HR diagram as a post-AGB star (late thermal pulse---LTP) or along the WD cooling track (very-LTP) causes the star to expand and become a `born again’ AGB star.  Merger remnants exploding while still on the AGB form {hydrogen-polluted} peculiar SNe Ia, while an explosion inside an inflated born-again star results in an early flux excess in the light curve of the SN Ia. The fraction of systems that might show an early flux excess due to LTP/VLTP is $< {\rm few} \times 10^{-4}$ of all SNe Ia, much below the observed fraction. In the frame of the CD scenario SNe Ia with early flux excess result from SN ejecta collision with planetary nebula fallback gas, or from mixing of $^{56}$Ni to the outer regions of the SN ejecta. Ongoing sky surveys might find about one case per year where LTP/VLTP influences the SN light curve.
}

 \authorrunning{N. Soker}            
\titlerunning{Pre-explosion helium shell flash in type Ia supernovae}  
   
      \maketitle
\section{INTRODUCTION}
\label{sec:intro}

There are several theoretical scenarios for the ignition of carbon oxygen white dwarfs (CO WDs) as type Ia supernovae (SN Ia; e.g., \citealt{Hoeflich2017, LivioMazzali2018, Soker2018Rev, Wang2018,  Jhaetal2019NatAs, RuizLapuente2019, Soker2019Rev, Ruiter2020} for less than five years old reviews). I classify these scenarios as follows (see \citealt{Soker2019Rev} for a comparison table of the six scenarios).

\begin{enumerate}
\item The \textit{core-degenerate (CD) scenario} involves the merger of a CO, or HeCO, WD with the core of a massive asymptotic giant branch (AGB) star. The merger remnant is a WD close to the Chandrasekhar mass limit that explodes at a later time. The time to explosion is the merger to explosion (MED) delay time $t_{\rm MED}$ (e.g., \citealt{KashiSoker2011, Ilkov2013, AznarSiguanetal2015}). The possibility that the CO WD merger remnant maintains a very small mass of helium and the possible consequences are the subjects of the present study. 
\item The \textit{double degenerate (DD) scenarios and the DD-MED scenario} involve the merger of two WDs (e.g., \citealt{Webbink1984, Iben1984}) that lose energy to gravitational waves. Helium might play a crucial role in igniting the carbon in the merging WDs (e.g., \citealt{YungelsonKuranov2017, Zenatietal2019, Peretsetal2019}). The explosion might take place during the merger itself (e.g., \citealt{Pakmoretal2011, Liuetal2016, Ablimitetal2016}), or at a MED time (e.g., \citealt{LorenAguilar2009, vanKerkwijk2010, Pakmor2013, Levanonetal2015, LevanonSoker2019, Neopaneetal2022}). 
\item In the \textit{double-detonation (DDet) scenario} ignition of helium detonation  that a CO-rich WD accretes from a companion { subsequently triggers a carbon core detonation that explodes the WD} (e.g., \citealt{Woosley1994, Livne1995, Shenetal2018}). 
\item In the \textit{single degenerate (SD) scenario} a CO WD accretes hydrogen-rich gas from a non-degenerate companion (e.g., \citealt{Whelan1973, Han2004, Orio2006, Wangetal2009, MengPodsiadlowski2018}). The WD might explode as soon as it approaches the Chandrasekhar-mass, or it might explode after a long delay (long MED) after losing angular momentum (e.g., \citealt{Piersantietal2003, DiStefanoetal2011, Justham2011}). 
\item  In the \textit{WD-WD collision (WWC) scenario} the two WDs explode as they collide at about their free fall velocity (e.g., \citealt{Raskinetal2009, Rosswogetal2009, Kushniretal2013, AznarSiguanetal2014}). Studies (e.g., \citealt{Toonenetal2018, HallakounMaoz2019, HamersThompson2019, GrishinPerets2022}) found that the WWC scenario contributes  $<1 \%$ of all SNe Ia.    
\end{enumerate}

The community is far from a consensus on the leading SN Ia scenario(s). Therefore, studies continue to explore each one of these stellar binary scenarios
(e.g., some examples from the last two years, \citealt{CuiXetal2020, Zouetal2020, Blondinetal2021, Clarketal2021,  Ablimit2021, Chandraetal2021, Liuetal2021, MengLuo2021, Michaely2021, WangQetal2021, Yamaguchietal2021, Zengetal2021, Ablimit2022, Acharovaetal2022, Chuetal2022, Cuietal2022, Dimitriadisetal2022, Ferrandetal2022, Lachetal2022, Liuetal2022, LivnehKaatz2022, Mazzalietal2022, Pakmoretal2022, Patraetal2022, Piersantietal2022, RauPan2022, RuizLapuenteetal2022, Sanoetal2022, SharonKushnir2022, Shingles2022}).

My view (see table in \citealt{Soker2019Rev}) is that the CD scenario and DD-MED scenario account for most normal SNe Ia. The main differences between the core-WD merger in the CD scenario and the WD-WD merger in the DD scenario are the hot core and the hydrogen-rich envelope at merger in the CD scenario. In the present study I propose that in some cases the core-WD merger remnant in the CD scenario might maintain sufficient amount of helium to experience  a late helium shell flash, i.e., a late thermal pulse, (LTP), or a very late thermal pulse (VLTP). 

In single star evolution a post-AGB star might experience a LTP, i.e., a helium shell flash while it is still burning hydrogen and moves leftward horizontally on the HR diagram, or it might experience a VLTP, i.e., a helium shall flash after hydrogen burning has ceased and the post-AGB is now on the WD cooling track.  
The parameters that determine whether a post-AGB star experiences a LTP, a VLTP, or none, are the mass of the helium layer when it leaves the AGB and where in the thermal pulse cycle on the AGB the star departs the AGB (e.g., \citealt{Iben1987, LawlorMacDonald2006}). The massive WD that we deal with here is formed from the merger of the core with a WD, and therefore it is not straightforward to deduce on its evolution by using single-star evolutionary tracks. I will rather use some general properties of single-star evolution to speculate on possible outcomes for the merger product. 

In section \ref{sec:Ingredients} I discuss the conditions for a pre-explosion LTP or VLTP. In section \ref{sec:Retaing001} I explore the way by which the core-WD merger remnant retains some helium, and in section \ref{sec:timescales} I refer to the relevant timescales of the scenario. In section \ref{sec:Rate} I estimate the rate of SNe Ia where LTP/VLTP might influence the light curve, and in section \ref{sec:Consequences} I discuss possible consequences. I discuss and summarize this somewhat speculative study in section \ref{sec:Summary}. 

\section{The conditions for a pre-explosion (very) late thermal pulse (LTP/VLTP)}
\label{sec:Ingredients}

The occurrence of a pre-explosion LTP/VLTP requires the following ingredients. It is the first condition (section \ref{subsec:RetainingHelium}) that I expect to be rare. The second condition (section \ref{subsec:DelayTime}) might hold for a large fraction of SNe Ia. 
 
\subsection{Retaining helium in the merger process}
\label{subsec:RetainingHelium}
In low-mass single-star evolution an LTP (when hydrogen still burns in a thin shell) or a VLTP (after the hydrogen burning has ceased) take place when the helium mass on top of the CO core when the star leaves the AGB is $\Delta M_{\rm He} \simeq 0.01 M_\odot$ (e.g., \citealt{Iben1987, LawlorMacDonald2006}). On the other hand, more massive stars, super-AGB stars that form oxygen-neon cores, have weak thermal pulses with short time intervals between pulses (e.g., \citealt{Siess2010}; see discussion of thermal pulses by, e.g., \citealt{GautschyAlthaus2007}). This implies that only a small hydrogen mass is required to lead to the next pulse. \cite{Siess2010} finds that for a core of about $1.3 M_\odot$ the hydrogen mass advances by only $5 \times 10^{-5} M_\odot$ between pulses. Two comments related to these values are as follows. (1) The young merger product is more extended than an isolated massive WD of the same mass \citep{Neopaneetal2022}, and so the required mass for a LTP/VLTP is larger than for a single star of the same mass. (2) The relevant point for this study is that if the LTP/VLTP is very weak, lasts for a short time, and does not contain much mass, then it has only a minor effect on the light curve of the SN Ia. 

Following the above discussion I will scale quantities with the requirement of massive helium shell, but note that the scaling might overestimate the mass required for the LTP/VLTP to take place, and it is possible that in all merger remnants the LTP/VLTP are very weak. In any case, I will conclude that strong LTP/VLTP must be very rare (less than few per cents) among all SNe Ia that result from the CD scenario.   

In the single-star simulations the mass of the post-AGB star that is evolving to become a WD is $\simeq 0.6-0.9 M_\odot$, and the WD evolves as a slowly rotating single star. In the scenario I study here the WD is a merger remnant that is rapidly rotating and its mass is $\simeq 1.4 M_\odot$. I take the condition on the remaining helium mass in the merger remnant to be 
\begin{equation}
\Delta M_{\rm He,TP} \approx 0.01 M_\odot; \qquad {\rm Condition~1}. 
\label{eq:MinimumHelium}    
\end{equation}
As stated above, because of the high mass of the merger remnant, this mass limit can be smaller even. 

Note that condition 1 (equation \ref{eq:MinimumHelium}) does not require that immediately after the merger process ends this is the helium mass. It rather allows also for an increase of the helium mass by hydrogen burning after the merger ends (section \ref{subsec:EnvelopeHelium}). 

\subsection{A merger to explosion delay (MED) time}
\label{subsec:DelayTime}
In single-star evolution typical LTP/VLTPs occur at about a thousand to tens of thousands of years after the star has left the AGB  (e.g.,  \citealt{Ibenetal1983, Blocker2001}). In the case of a massive WD that is a remnant of a core-WD merger the time scale can be shorter even. I will return to this timescale in section \ref{sec:timescales}. Here I only point to the requirement of a MED. 

The requirement that, at least some, progenitors of SNe Ia have a MED, i.e., a time delay from merger (or from a mass transfer event) to explosion comes from the observational findings that many SNe Ia have no close circumstellar matter (CSM; see review by \citealt{Soker2018Rev}) at explosion time, and that many SNe Ia have a more or less spherical explosion (see review by \citealt{Soker2019Rev}). 

Studies have considered a MED of tens of years (e.g., \citealt{Neopaneetal2022} for a recent paper on the DD scenario) to tens of thousands of years (e.g., \citealt{Soker2022Delay} for a recent paper on the CD scenario). 
For the VLTP to have any role on the light curve of the SN, the LTP/VLTP cannot occur a long time before the explosion. The typical time of the LTP/VLTP after the AGB phase sets an upper limit on the MED of $\approx 10^5 \yr$. 
Overall, for a pre-explosion LTP/VLTP to occur and play a role I require the MED time to be 
\begin{equation}
t_{\rm med} \la 10^5 \yr; \qquad {\rm Condition~2}. 
\label{eq:MEDconstraints}    
\end{equation}

Condition~2 (equation \ref{eq:MEDconstraints}) does not impose a strong limit on pre-explosion LTP/VLTP if we consider that, crudely, $\approx 50 \%$ of SNe Ia might obey this condition \citep{Soker2022Delay}. It is Condition~1 (equation \ref{eq:MinimumHelium}) that strongly constrains the number of pre-explosion LTP/VLTP. I therefore turn to study the possibility of the merger remnant to retain some helium.

\section{Retaining low helium mass}
\label{sec:Retaing001}

\subsection{The CD scenario versus the DD scenario}
\label{subsec:Retaing}

Critical to the occurrence of a LTP/VLTP is that some helium survives the merger process. 
When two WDs merge in the DD scenario the helium, which one or two of them have, easily burns because of the high temperatures that the outer zones achieve (e.g., \citealt{Danetal2012, Pakmoretal2013, Peretsetal2019, Pakmoretal2021}). Therefore, it is not clear at all whether any helium can survive the merger of two WDs. The answer will come from future 3D simulations. 

I therefore consider only the CD scenario where the merger takes place at the end of the CEE \citep{KashiSoker2011}. In this case the helium on the outer zone of the core is already very hot, $\simeq 10^8 \K$. It is possible therefore that this helium is not compressed to high enough densities to be ignited. On the other hand, the violent merger process likely mixes helium with carbon, a mixture that burns more easily.  Again, only 3D numerical simulations can determine whether some helium from the core survives the core-WD merger in the CD scenario. Therefore, I first consider the case where hydrogen and helium from the envelope survive. 

\subsection{Surviving envelope helium}
\label{subsec:EnvelopeHelium}

To have a core-WD merger in the CD scenario the mass of the envelope of the AGB star at the time it engulfs the WD secondary should be $M_{\rm AGB,env} \ga 3 M_\odot$ (e.g., \citealt{Sokeretal2013, Canalsetal2018}). 

I use the Reimers mass loss rate (\citealt{Reimers1975}) in the form $\dot M_{\rm wind} \simeq 3 \times 10^{-13} LM^{-1} R M_\odot \yr^{-1}$, where the stellar luminosity $L$, mass $M$, and radius $R$ are in solar units. I note that other expressions are also possible (e.g., \citealt{SchroderCuntz2007}). For this mass loss rate, which is linear with the stellar luminosity, the ratio of the rate at which envelope burning adds helium to the core, $\dot M_{\rm He}$, to the wind mass loss rate, $\dot M_{\rm wind}$, is 
\begin{equation}
\frac{\dot M_{\rm He}}{\dot M_{\rm wind}} \simeq 0.3 
\left( \frac{M_{\rm rem}}{1.4 M_\odot} \right)
\left( \frac{R_{\rm g}}{200 R_\odot} \right)^{-1},
\label{eq:MassLosesRates}
\end{equation}
where $M_{\rm rem}$ is the mass of the merger remnant, and for the giant radius $R_{\rm g}$ I take a moderate radius because the envelope mass is low.
To retain a helium mass that allows a LTP/VLTP, i.e., $\Delta M_{\rm He,TP} \approx 0.01 M_\odot$ (equation \ref{eq:MinimumHelium}), the merger process should leave an envelope mass of 
\begin{equation}
M_{\rm rem, env} \approx 0.03 M_\odot \simeq 0.01 M_{\rm AGB,env}. 
\label{eq:EnvelopeMass}
\end{equation}

\cite{KashiSoker2011} estimated that to facilitate a core-WD merger in the CD scenario a fraction of $\eta_{\rm CBD} \simeq 0.01-0.1$ of the common envelope remains bound in a circumbinary disk at the end of the CEE. 
Here I require that either in addition to the circumbinary disk there is a low-mass envelope that survives the CEE, or that a fraction of the circumbinary disk remains bound after merger and it is inflated to form a low-mass envelope. 

The conclusion is that to have a LTP/VLTP the merging process should leave about one per cent or more of the initial AGB envelope mass bound to the merger remnant. I expect that only a minority, if at all, of SN Ia progenitors in the CD scenario will leave such an envelope, but that nonetheless it is possible. 

\subsection{Surviving core helium}
\label{subsec:CoreHelium}

I here estimate that not enough helium to lead to a LTP/VLTP from the core survives the merger. 

As I mentioned above (section \ref{subsec:Retaing}) it is not clear at all that the merger process leaves any helium from the core. In the core-WD merger process either the WD or the core stay intact, and the other object is tidally destroyed. The helium, being on the outer part of the core, is mixed with the CO of the destroyed object, being the WD or the core. Therefore, if the process leaves any helium, the helium is mixed with the carbon and oxygen in the outer parts of the merger remnants. At the outer regions the densities are too low for the pycnonuclear reactions that \cite{Chiosietal2015} studied to take place. \cite{Chiosietal2015} proposed an isolated sub-Chandrasekhar WD scenario for SNe Ia. In their scenario pycnonuclear reactions between carbon and tiny amounts of hydrogen or helium during the crystallisation of the WD { cause a combination of WD temperature and density that is sufficient to initiate carbon burning, i.e., $^{12}{\rm C}+^{12}{\rm C}$ and up to $^{56}$Ni. It is the carbon burning that explodes the WD. }  

If some helium is mixed with the CO in the outer zone of the WD it diffuses out. I very crudely estimate the diffusion coefficient of helium through CO matter based on expression from, e.g., \cite{Paquetteetal1986} and \cite{Caplanetal2022}, to be $D \approx 0.1 (T/10^8 \K)^{3/2} (\rho/10^6 \g \cm^{-1})^{-2/3} \cm^2 \s^{-1}$. 
The helium is spread in a layer of width $\Delta r \approx 1000 \km$. The diffusion time will be therefore $\tau_{\rm diff} \approx (\Delta r)^2 /D$.
For the above parameters, and since the WD cools over this long time scale, I estimate that $\tau_{\rm diff} > 10^9 \yr$. (I get the same crude timescales from scaling the tables that \citealt{Koester2009} give).  
By that time the WD has cooled. I conclude that if the helium mixture with carbon and oxygen in the hot WD merger remnant did not ignite the helium (but I do expect ignition of helium), the floating of helium on the surface of the WD is unlikely to ignite a VLTP. 
    
\section{Relevant timescales}
\label{sec:timescales}

In single-star evolution a LTP (a post-AGB helium shell flash while hydrogen still burns) takes place hundreds to thousands of years after departure from the AGB (e.g., \citealt{Blocker2001}). For a VLTP (when hydrogen burning has already ceased) the star needs to fade first to a luminosity of $L_{\rm VLTP} \approx {\rm few} \times 100 L_\odot$ along the WD cooling track (e.g.,  \citealt{Ibenetal1983, Blocker2001}). This implies that VLTPs take place tens of thousands of years after departure from the AGB, with longer timescales for more massive stars (e.g., \citealt{BlockerSchonberner1990, Bloecker1995}, for evolutionary times on the WD cooling track). Overall, I crudely take the LTP/VLTP of the merger remnant to occur sometime in the post-AGB age of 
\begin{equation}
100 \yr \la t_{\rm AGB-TP} \la 10^5 \yr.
\label{eq:LTP/VLTP}    
\end{equation}
This is the reason for condition 2 (equation \ref{eq:MEDconstraints}).
I do note that the merger remnant is expected to be highly magnetized (e.g., \citealt{Pelisolietal2022}), { and that magnetic fields might play a role in SN Ia progenitors in different scenarios (e.g., \citealt{AblimitMaeda2019}). }. Future studies should examine the role of the high magnetic fields in the LTP/VLTP. 

For a surviving envelope mass of $M_{\rm rem, env}$, and using the same mass loss rate as in section \ref{subsec:EnvelopeHelium} and for a remnant mass of $M_{\rm rem}=1.4 M_\odot$, the time from the merger event to the termination of the AGB by the merger remnant is 
\begin{equation}
\begin{split}
& \Delta t_{\rm mer-AGB}  \approx 1.1 \times 10^4  
\left( \frac{M_{\rm rem, env}}{0.03 M_\odot} \right) 
\\ & \times
\left( \frac{L_{\rm rem}}{5 \times 10^4 L_\odot} \right)^{-1}  
\left( \frac{R_{\rm g}}{200 R_\odot} \right)^{-1} \yr,
\label{eq:t(remAGB)}
\end{split}
\end{equation}

Another process that determines the post-merger evolution is angular momentum loss through the magnetic dipole radiation torque. This timescale might set the MED time in the CD scenario (e.g., \citealt{IlkovSoker2012}) and in the DD scenario (e.g., \citealt{Neopaneetal2022}).  Several parameters determine the angular momentum loss timescale. Some of them are more secure, like the moment of inertia of the WD. Assuming that the merger remnant is rapidly rotating and has a mass of $\simeq 1.4 M_\odot$, there are two undetermined parameters that influence the angular momentum loss timescale the most. These are the dipole magnetic field of the remnant $B$, and 
the angle between the magnetic axis and the rotation axis of the WD remnant, $\delta$. The angular momentum loss timescale dependence on these two parameters is (for more details see \citealt{IlkovSoker2012} and \citealt{Neopaneetal2022})
\begin{equation}
\begin{split}
\tau_{\rm B} \approx   10^{4}
\left(\frac{B}{10^9 \G}\right)^{-2}
\sin^2 \delta \yr. 
\end{split}
\label{eq:taub}
\end{equation}
Note that \cite{IlkovSoker2012} scale (their equation 10) by $B=10^{8} \G$ and by $\sin \delta = 0.1$, while \cite{Neopaneetal2022} scale (their equation 10) by $B=10^{10} \G$ and by $\sin \delta = 1$. 
If the angular momentum loss timescale determines the time to the WD explosion then $t_{\rm MED} \simeq \tau_{\rm B}$. 

If indeed condition 2 (equation \ref{eq:MEDconstraints}) holds, then the explosion takes place before the nebula that the CEE (that leads to the merger) ejected disperses to the ISM, although it might be at a large distance from the explosion. This nebula is a planetary nebula, as the WD remnant ionises the nebula. The explosion of a SN inside a planetary nebula is termed SNIP. In \citep{Soker2022Delay} I crudely deduced by analysing observations that SNIPs account for $\approx 50 \%$ of all normal SNe Ia (i.e., not including peculiar SNe Ia). 

\section{The rate of the rare LPT/VLTP SNIPs}
\label{sec:Rate}
 
Equation (\ref{eq:t(remAGB)}) shows that if the merger remnant maintains enough envelope mass to experience a strong LTP/VLTP, then its envelope mass dictates that it  stays an AGB star for $\Delta _{\rm mer-AGB} \approx 10^4 \yr$ after merger. 
In \cite{Soker2022Delay} I crudely estimated that about half of all SNe Ia are SNIPs that take place within a time scale of $t_{\rm SNIP} \approx 10^4 \yr$ after the end of the CEE of the WD and the AGB star. The equality $\Delta _{\rm mer-AGB} \approx t_{\rm SNIP}$ implies that about half the merger remnants that maintain a large enough envelope mass to experience a helium shell flash will explode while the merger remnants is still an AGB star. This holds whether the helium shell flash ignites the { the carbon in the } core or not. { The carbon detonation explodes the star. } The ignition might result from a last helium shell flash during this final AGB phase that ignites the carbon. Else, the loss of angular momentum might lead to the core explosion (see section \ref{subsec:Triggering}). 

The outcome of an explosion of the core while there is a hydrogen-rich envelope of radius $\approx 100 R_\odot$ and a mass of $\approx 0.01 M_\odot$ is a peculiar SN Ia if the hydrogen is not detected. It is peculiar because the relatively massive envelope, which in turn might lead to an early flux excess much higher than in normal SNe Ia with early flux excess (section \ref{subsec:LightExcess}). { If the hydrogen is detected, then it might at first be classified as a peculiar SN II. However, later analysis of the light curve will reveal extremely low hydrogen mass alongside a $^{56}$Ni mass that is much larger than what typical SNe II have. This will point to a hydrogen-polluted peculiar SN Ia.  }

Because I am not aware of such peculiar SNe Ia and SN II, I conclude that the number of cases, $N_{\rm TP,SN}$, where the WD-core merger remnant maintains a large enough envelope to stay an AGB and experience a helium shell flash on the AGB (thermal pulse) or experience a LTP/VLTP are very rare, at most few per cents of the number of all SNe Ia, $N_{\rm Is}$. Namely, 
\begin{equation}
N_{\rm TP,SN} < {\rm few} \times 0.01 N_{\rm Ia}. 
\label{eq:fraction}
\end{equation}
If and when such peculiar SNe are identified, we might be able to better constrain this fraction. 

I emphasise again that the mass limit for LTP/VLTP of the massive merger remnant might be much smaller than what condition 1 (equation \ref{eq:MinimumHelium}) gives. In that case the merger remnant stays a short time on the AGB, and even if explosion does take place during the AGB phase the effect of the very low mass envelope is small. Therefore, weak LTP/VLTP, those that result from a very low helium mass of $\Delta M_{\rm He,TP}  \ll 10^{-2} M_\odot$, might be more common. This should be explored in future studies. 

\section{Possible consequences of LTP/VLTP}
\label{sec:Consequences}

I speculate here on three possible consequences of a LTP/VLTP. As I  stated in section \ref{sec:timescales}, because all explosions occur at a short time ($\la 10^5 \yr$) after the AGB phase for the LTP/VLTP to play a role, all these SNe Ia are SNIPs, i.e., SNe Ia inside planetary nebulae. 

\subsection{Triggering WD remnant explosion}
\label{subsec:Triggering}

In single star evolution the WD mass does not reach a mass larger than about $1.1 M_\odot$. Calculations show that a LTP/VLTP does not ignite the WD (otherwise we would have a much larger number of SNe Ia than observed). { By `igniting the WD' I refer to the ignition of carbon burning in the core that it turns explodes the WD. } However, a LTP/VLTP on a merger remnant that is about the Chandrasekhar mass, $M_{\rm rem} \simeq 1.4 M_\odot$, might ignite the WD, similar to the ignition by a thin helium shell in some channels of the DDet scenario (e.g., \citealt{ShenMoore2014}). \cite{ShenMoore2014} find that a pure helium shell mass of $\simeq {\rm few} \times 0.001 M_\odot$ can ignite a WD of mass $\ga 1.3 M_\odot$. A small mixture of CO into the helium ease the ignition of helium. Since I take the helium mass to be $\Delta M_{\rm He,TP} \ga 0.01 M_\odot$, which is condition 1 (equation \ref{eq:MinimumHelium}), the LTP/VLTP might ignite the WD. 
During the LTP/VLTP the star is hydrogen-deficient with total hydrogen mass of $\la 10^{-5} M_\odot$ (e.g., \citealt{Herwigetal1999}), and so the explosion will be indeed classified as a SN Ia (or a peculiar SN Ia). 

Previous studies of the CD scenario (e.g., \citealt{IlkovSoker2012}) assume that the loss of angular momentum leads to the explosion of the Chandrasekhar-mass merger remnant, with a time scale as given by equation (\ref{eq:taub}). In addition to this possibility, I raise here the  possibility that in rare cases (equation \ref{eq:fraction}) LTP/VLTP ignites the WD.  

Chandrasekhar-mass WDs that explode in the SD scenario by detonation alone, without deflagration to detonation transition, produce too much $^{56}$Ni to be compatible with observations of normal SNe Ia. However, there are some super-Chandrasekhar-mass WDs that do have larger than usual  $^{56}$Ni masses (e.g., \citealt{Debetal2022, Dimitriadisetal2022} for recent studies). 
Moreover, in a recent study \cite{Neopaneetal2022} argue that a Chandrasekhar-mass merger product that explodes within a short time after merger, namely, a short MED time, might have a different density structure than a Chandrasekhar WD in the SD scenario (be more extended), and therefore has different properties at explosion. Specifically, \cite{Neopaneetal2022} find that their calculated synthetic spectra of a merger product that explodes by detonation (with no deflagration to detonation transition) with $t_{\rm MED}= \simeq 100 \yr$ in the DD scenario matches normal SN spectra. 
Future studies should examine the properties of detonation of Chandrasekhar-mass WDs with MED times of up to $t_{\rm EMD} \simeq 10^4 - 10^5 \yr$, and determine whether they produce normal or peculiar SNe Ia.

Yet another possibility is that a LTP/VLTP ignites sub-Chandrasekhar WD remnants, but these are more massive than what single-star evolution gives. \cite{Royetal2022} find that helium detonation does not ignite a WD of $\simeq 1 M_\odot$ in the DDet scenario. They further suggest that helium detonation might ignite the most massive CO WDs. 
Namely, LTP/LTP events, which are similar in many respects to helium detonation in the DDet scenario, might ignite WD merger remnants in the mass range of
\begin{equation}
1.2 M_\odot \la    M_{\rm WD,mer}({\rm LTP/VLTP}) < 1.4 M_\odot .  
\label{eq:MWDmer}
\end{equation}
Even in this case the WD needs to be more extended than a cold WD (as \citealt{Neopaneetal2022} find) to be a normal SN Ia, because cold WDs that explode in the DDet scenario have to be of mass $\simeq 1 M_\odot$ to account for normal SNe Ia (e.g., \citealt{Shenetal2018}). 

\subsection{WD explosion inside a born-again AGB star}
\label{subsec:BornAgain}

Different simulations show that the born-again star can stay inflated, i.e., $R \gg 1 R_\odot$ (e.g., \citealt{Schoenberner1979}) for hundreds of years (e.g., 
\citealt{Althausetal2005, Guerreroetal2018, Lawlor2021}).
However, more massive stars at the time of LTP/VLTP evolve faster (e.g., \citealt{Bloecker1995, BloeckerSchoenberner1997}). On top of these, the time of evolution depends on the uncertain efficiency of convective mixing of elements (e.g., \citealt{Herwig2002, Althausetal2005}), and the evolution time seems to be short (e.g., \citealt{Herwig2002, LechnerKimeswenger2004}), down to several years (e.g., \citealt{Claytonetal2006}).  Overall, I scale the WD-inflated phase after a LTP/VLTP with a time scale of $\tau_{BA} \approx 100 \yr$, but it might be shorter even. This implies that only $\la 1 \%$ of the SNe Ia that explode within $10^4 - 10^5 \yr$ from the termination of the AGB will take place inside a born-again envelope. With equation (\ref{eq:fraction}) I conclude that the fraction of SN Ia inside an inflated born-again envelope is extremely low 
\begin{equation}
N_{\rm BA,SN} < {\rm few} \times 10^{-4} N_{\rm Ia}. 
\label{eq:fractionBA}
\end{equation}

The gas that the born-again AGB star ejects might be H-poor and He-poor (e.g., \citealt{MontoroMolinaetal2022}). The inflated envelope of the born-again star in these cases is H-poor and He-poor. However, calculations show that the inflated envelope can be H-rich and He-rich as well (e.g., \citealt{Althausetal2007}). 
The envelope can be also He-rich and H-poor, as with  V605~Aql, the central star of the planetary nebula A58  (e.g., \citealt{Claytonetal2006}).

An explosion inside the very low mass, $\approx 10^{-3} M_\odot$, H-poor or H/He-poor envelope would still be classified as a SN Ia. However, the very hot thin envelope will cool in the first hours to days to give an early flux excess, similar to that when the ejecta collides with circumstellar matter (section \ref{subsec:LightExcess}). The outcome is therefore a SN Ia with early flux excess in the frame of the CD scenario.    

\subsection{Explosion into a born-again ejecta}
\label{subsec:LightExcess}

A born-again star, which following a LTP/VLTP becomes an AGB star again, ejects a mass of $\approx 10^{-4}-10^{-3}$ (but the mass can be larger even) in a clumpy morphology (e.g., \citealt{Jacobyetal2020} for a discussion and references). 
\cite{Fangetal2014} model the born-again phase with a duration of 20 years, a wind velocity of $20 \km \s^{-1}$, and a mass loss rate of $10^{-4} M_\odot \yr^{-1}$.  
Following these studies and others, I take a typical ejecta of a born-again star to be clumpy (e.g., \citealt{Kerberetal2009, Hinkleetal2020, RodriguezGonzalezetal2022}), to contain a mass of $\approx 10^{-4} - 10^{-3} M_\odot$, and to expand at ${\rm few} \times 10 \km \s^{-1}$. However, for the inner and slowest parts of the ejecta I take a velocity of $\simeq 10 \km \s^{-1}$. 

The born-again ejecta adds up to the planetary nebula shell and does not play a prominent additional role at late times. The born-again ejecta can lead to an early flux excess of the SN Ia only if the SN ejecta collides with the born-again ejecta in the first few days, before the SN Ia becomes bright by radioactive decay.
The process is like the early flux excess that might result from the collision of the SN Ia ejecta with the circumstellar matter in some other SN Ia scenarios (e.g., \citealt{RaskinKasen2013, Levanonetal2015, Kromeretal2016, PiroMorozova2016}).  

For the fastest supernova ejecta at $v_{SN} \simeq 20,000 \km \s^{-1}$ to collide with the slow born-again ejecta moving at $10 \km \s^{-1}$ within three days after explosion, the SN explosion should take place with a time delay after the LTP/VLTP of only $\Delta t_{TP-SN} \la 16 \yr$. The fraction of such systems that lead to early flux excess (ELE) is less even than the fraction of SNe Ia inside a born-again envelope, and it is only 
\begin{equation}
N_{\rm ELE,SN} < {\rm few} \times 10^{-5} N_{\rm Ia}. 
\label{eq:fractionELE}
\end{equation}

Overall, the fraction of systems that explode inside an extended born-again envelope or a decade after the star contracts back, is about $\la 1 \%$ of all cases with LTP/VLTP. These SNe Ia will have an early flux excess. But equations (\ref{eq:fractionBA}) and (\ref{eq:fractionELE}) imply that such systems can account for no more than a fraction of ${\rm few} \times 10^{-4}$ of all SNe Ia. 

\section{Discussion and Summary}
\label{sec:Summary}

In this study I raise the (speculative) possibility that within the frame of the CD scenario (section \ref{sec:intro}) in rare cases the merger process of the core and the WD maintains an envelope mass (equation \ref{eq:EnvelopeMass}) that is sufficiently massive to cause a later helium shell flash. The helium shell flash might take place while the merger remnant is still on the AGB (section \ref{sec:Rate}), while the star moves to the left on the HR diagram as a post-AGB star (this is a LTP), or along the WD cooling tack (VLTP). 

Although the merger remnant is massive, it is hot and more extended than a WD of the same mass that evolves as a single-star or grows by slow mass accretion \citep{Neopaneetal2022}. In addition the merger remnant is rapidly rotating and sustains a strong magnetic field. Future studies should determine the helium mass that can drive LTP/VLTP on such complicated WD-core merger remnants. In this study I scaled the required helium mass for an LTP/VLTP as in equation (\ref{eq:MinimumHelium}), but one should keep in mind that the required mass might be much smaller.   

The sufficiently massive envelope according to condition 1 (equation \ref{eq:MinimumHelium}) implies that the merger remnant spends $\approx 10^4 \yr$ on the AGB (equation \ref{eq:t(remAGB)}).
I estimated that in the frame of the CD scenario the fraction of merger remnants that maintain sufficiently massive envelope to have a LTP/VLTP is at most few per cents of all SNe Ia (equation \ref{eq:fraction}). Otherwise there will be a too large number of WD explosions inside a thin AGB envelope. 
However, if the required helium mass is smaller because of the massive remnant, i.e., $\Delta M_{\rm He,TP}  \ll 10^{-2} M_\odot$ (see discussion in section \ref{subsec:RetainingHelium} and \ref{sec:Rate}), then the merger remnant stays on the AGB for a much shorter time and the effect of the very low mass envelope on the SN Ia light curve is small. Therefore, weak LTP/VLTP that results from these low mass helium shells might be more common.
  
In cases the WD merger remnant explodes while the merger remnant is still on the AGB the outcome is a peculiar SN Ia if no hydrogen or helium are detected, { or a speculative peculiar hydrogen-polluted SN Ia if hydrogen is detected (section \ref{sec:Rate}). }
If the WD merger-remnant explodes inside an inflated born-again star, or within $<20 \yr$ after the born-again star has contracted back, then the outcome is an early flux excess in the light curve of the SN Ia. The fraction of systems that might show an early flux excess due to this process is the combined rates of equations (\ref{eq:fractionBA}) and (\ref{eq:fractionELE}), a fraction of  $< {\rm few} \times 10^{-4}$ of all SNe Ia. 
 
In recent studies \cite{Deckersetal2022} estimate that SNe Ia with early flux excess, such as SN~2015bq \citep{LiZhangDaietal2022}, amount to $\simeq 18 \pm 11 \%$ of all SNe Ia at redshift of $z<0.07$, and \cite{Mageeetal2022} estimate this fraction to be $\simeq 28 \pm 11 \%$. 
My conclusion is that the inflated envelope and mass ejection by LTP/VLTP events might at most account for a very small fraction of SNe Ia with early flux excess.

\cite{Deckersetal2022} also find that SNe Ia with early flux excess originate from younger stellar populations. This, and the early flux excess in some super-Chandrasekhar SNe Ia (e.g., SN~2020hvf;  \citealt{Jiangetal2021}), might suggest that some early flux excess SNe Ia come from the CD scenario. The CD scenario can account for super-Chandrasekhar SNe Ia and can explain a short CEE to explosion delay time ($t_{\rm CEED}$; \citealt{Soker2022Delay}). Therefore, it is possible that the CD scenario might account for early flux excess in some SNe Ia, but not by LTP/VLTP. Either the early flux excess results from ejecta-CSM collision like in other SN Ia scenarios (e.g., \citealt{RaskinKasen2013, Levanonetal2015, Kromeretal2016, PiroMorozova2016}), or the explosion mixes $^{56}$Ni to the outer regions of the SN ejecta as in some other SN Ia scenarios (e.g., \citealt{PiroMorozova2016, Jiangetal2018}). The CSM might be gas from the planetary nebula that falls back towards the remnant. The fallback process requires further study. { In a very recent paper \cite{Ashalletal2022} suggest that in some cases Doppler shift of high velocity gas can account for blue flux excess. This process is relevant also to the CD scenario. }

The helium burning during the core-WD merger in the CD scenario, and in rare cases in an LTP/VLTP, nucleosynthesises the radioactive isotope $^{44}$Ti, like helium burning in the DDet scenario (e.g., \citealt{Royetal2022}). In most SNe Ia according to the CD scenario, as I found above, LTP/VLTP does not take place. As well, most explosions take place more than a thousand years after merger. Because the radioactive isotope $^{44}$Ti has a half-life of 60 years I do not expect the presence of more $^{44}$Ti than what the explosion by deflagration to detonation transition predicts (see comparison between models by, e.g, \citealt{Trojaetal2014}).
Recent studies suggest that most SNe Ia contain small masses of $^{44}$Ti, as the deflagration to detonation transition predicts or less (e.g., \citealt{Lopezetal2015,Weinbergeretal2020}). 
In rare cases the explosion takes place hundreds of years after merger. Then there might be some extra $^{44}$Ti. In these cases there is a close CSM because of the short delay from the CEE to explosion (short $t_{\rm CEED}$).
In the very rare cases of LTP/VLTP that triggers a SN Ia or take place a short time before the SN Ia explosion, I expect to find a somewhat larger $^{44}$Ti mass than what the deflagration to detonation transition predicts. 

Despite the very low fraction of SNe Ia with pre-explosion LTP/VLTP, the large number of SNe Ia (thousands per year) that ongoing sky surveys will find in coming years might lead to about one case per year where LTP/VLTP might have an influence on the light curve. 

\section*{Acknowledgments}

I thank Robert Fisher for enlightenment discussions and Marcelo M. Miller Bertolami, Yossef Zenati {and an anonymous referee} for useful comments. 
This research was supported by a grant from the Israel Science Foundation (769/20).



\end{document}